# Second quantization of a covariant relativistic spacetime string in Steuckelberg-Horwitz-Piron theory

Michael Suleymanov[1,2], Lawrence Horwitz[1,3,4] and Asher Yahalom[2]

[1]Department of Physics, Ariel University, Ariel, Israel

[2]Department of Electrical & Electronic Engineering, Ariel University, Ariel, Israel

[3]School of Physics, Tel Aviv University, Ramat Aviv, Israel

[4]Department of Physics, Bar Ilan University, Ramat Gan, Israel

## Abstract

A relativistic 4D string is described in the framework of the covariant quantum theory first introduced by Stueckelberg (1941) [1], and further developed by Horwitz and Piron (1973) [2], and discussed at length in the book of Horwitz (2015) [3]. We describe the space-time string using the solutions of relativistic harmonic oscillator [4]. We first study the problem of the discrete string, both classically and quantum mechanically, and then turn to a study of the continuum limit, which contains a basically new formalism for the quantization of an extended system. The mass and energy spectrum are derived. Some comparison is made with known string models.

## 1. Introduction to covariant relativistic dynamics

In classical Newtonian mechanics and in non-relativistic quantum mechanics (NRQM), the time, $t$, is a universal parameter of evolution that is the same in all frames (Galilean transformations) up to a constant which signifies the origin of the temporal axis. However, when we generalize our statement of the physical laws according to the theory of special relativity, the time $t$ is mixed with space coordinates by Lorentz transformations, and therefore, can no longer be taken as an evolution parameter. It has to be treated as an additional coordinate. Since the Schrödinger equation is not covariant, it is necessary to define a covariant quantum equation. The Klein-Gordon and Dirac equations are covariant, but there are serious problems in interpreting the solutions in a quantum mechanical framework. In the Klein-Gordon theory, the time component of the current, which should correspond to a density, is not positive definite. The solutions of the Dirac equation have a positive density, but both of these formulations were shown by Newton and Wigner (1949) [5] to be associated with wave functions that are non-local, i.e., interpreting them as quantum mechanical representations of the states of a particle, they have the property that for a particle at a definite determined point, the corresponding wave function has a spatial extent of the order of 1/m, where m is the mass of the particle. Therefore, there would be nonvanishing probability in regions where the particle is known to be not present with certainty, and the computation of interference effects would be unreliable.

In 1941, Stueckelberg presented a manifestly covariant classical and quantum theory, describing particle-antiparticle creation and annihilation, and introduced a new evolution parameter $\tau$. In this theory, energy and momentum are taken as independent quantities, which means that the particle mass becomes an observable and can have any value, defined by the state of the system. Horwitz and Piron (1973) [2] extended Stueckelberg's idea to deal with many body systems by asserting that the parameter $\tau$ of Stueckelberg is universal, as for the original notion of time introduced by Newton (this approach, from now on, will be called SHP theory) [3]. Using this covariant quantum theory, Horwitz and Arshansky (1989) [4] solved a family of two-body interaction models with action at a distance potentials (as functions of spacelike separation), including the relativistic (4D) harmonic oscillator with results in agreement with the nonrelativistic Schrödinger equation with relativistic corrections. This was solved taking the solutions of the differential equations in the submanifold called the restricted Minkowski space (RMS), as was discussed by Zmuidzinas in his paper on unitary representations of the Lorentz group (1965) [6].

This work studies a discrete spacetime string defined by two intrinsic parameters, "mass density" and tension. We diagonalize the Hamiltonian by taking the 4D Fourier transformation, giving a sum of independent harmonic oscillators in the space of the string's modes, therefore the solutions of Arshansky and Horwitz [4] can be used to describe the wave functions and spectrum of the string with the help of some simplifying assumptions. We then consider the limiting case of a continuous string.

## 1.1. Covariant relativistic classical mechanics

We assume the existence of a manifestly covariant Hamiltonian $K$, as a function of space-time coordinates $x^\mu = (ct, \vec{x})$, and four-momentum $p^\mu = (E/c, \vec{p})$, [we use the metric $(-+++)$] which satisfies the Hamilton type equations

$$\frac{dx^\mu}{d\tau} = \frac{\partial K}{\partial p_\mu} \qquad \frac{dp^\mu}{d\tau} = -\frac{\partial K}{\partial x_\mu} \qquad (1)$$

These are the equations of motion (EOM) of the event $x^\mu$ (a point in space time with properties $p^\mu$ and a scale parameter $M$). A classical Poisson bracket is therefore available. The evolution of the event, as function of $\tau$, traces out the world line.

For the free particle, one may choose [1]

$$K_0 = \frac{p^\mu p_\mu}{2M} = \frac{\vec{p}^2 - E^2/c^2}{2M} \qquad (2)$$

where $M$ is an intrinsic property of the particle with the dimensions of mass.

According to the EOM, the function $K$ is a constant of motion if it doesn't depend explicitly on $\tau$:

$$\frac{dK}{d\tau} = \frac{\partial K}{\partial x^\mu}\frac{dx^\mu}{d\tau} + \frac{\partial K}{\partial p^\mu}\frac{dp^\mu}{d\tau} = \frac{\partial K}{\partial x^\mu}\frac{\partial K}{\partial p^\mu} - \frac{\partial K}{\partial p^\mu}\frac{\partial K}{\partial x^\mu} = 0 \qquad (3)$$

The conserved quantity $K_0$ for the free particle corresponds to

$$K_0 = \frac{1}{2M}\left(-\left(\frac{E}{c}\right)^2 + \vec{p}^2\right) = -\frac{m^2 c^2}{2M} \qquad (4)$$

where $m$ is the observed mass. It follows from the equations of motion that for $ds^2 = -dx^\mu dx_\mu$ $(ds/d\tau)^2 = (m/M)^2$, so that $\tau$ coincides with the proper time when $m = M$ [1]. We call this the "mass shell" condition, in general, not preserved since $m$ is a dynamical variable. For a particle for which $p^\mu p_\mu \equiv -m^2 = -M^2$, $K$ has the value $-Mc^2/2$. One can show that in the nonrelativistic limit [7], the mass parameter $M$ can be considered as the Galilean target mass of the particle. Let us consider a spacetime dependent interaction [1] represented by a potential function,

---

[1] Aharonovich and Horwitz [14] have shown that for a charged scalar particle moving in its self-reaction electromagnetic field, m² may rapidly approach its mass shell value. Burakovsky, Horwitz and Schieve [15] have furthermore shown that for a boson gas in equilibrium there can be a high temperature phase transition (boson condensation, as in the work of Haber and Weldon [12]) for which m approaches M. It has, furthermore, been shown [7] that M corresponds to the Galilean limit (sharp) of the variable m.

$$K = \frac{p_\mu p^\mu}{2M} + V(x) \tag{5}$$

The covariant Hamiltonian for a many-body theory can be written as,

$$K = \sum_{i=1}^{N} \frac{p_{i\mu} p_i^\mu}{2M_i} + V(x_1,...,x_N) \tag{6}$$

and the EOM are then

$$\dot{x}_i^\mu \equiv \frac{dx_i^\mu}{d\tau} = \frac{\partial K}{\partial p_{i\mu}} \qquad \dot{p}_i^\mu \equiv \frac{dp_i^\mu}{d\tau} = -\frac{\partial K}{\partial x_{i\mu}} \tag{7}$$

Using the EOM, we see that

$$\frac{dx_i^\mu}{d\tau} = \frac{p_i^\mu}{M_i},$$

We can then write $K$ in terms of four-velocities,

$$K = \sum_{i=1}^{N} \frac{1}{2} M_i \dot{x}_{i\mu} \dot{x}_i^\mu + V(x_1,...,x_N) \tag{8}$$

Writing the time and space components separately, (1) may be written as,

$$\frac{dt}{d\tau} = \frac{E}{Mc^2} \tag{9}$$

$$\frac{d\vec{x}}{d\tau} = \frac{\vec{p}}{M} \tag{10}$$

and gives the known result

$$\frac{d\vec{x}}{dt} = \frac{\vec{p}}{E} c^2 \tag{11}$$

The coordinates, $t$ and $\vec{x}$, are the outcome of measurements in some frame (say, the laboratory), considered as a detection of signals (ignoring the propagation of the signal) generated by the particle (at time $\tau$), where, $t$ may be interpreted as a particle's clock rate (time) as it is measured in the laboratory.

Since $\frac{\partial K}{\partial \tau} = 0$, it can be chosen to be $K = \frac{-Mc^2}{2}$, hence, using (5),

$$-\frac{Mc^2}{2} = \frac{-m^2 c^2}{2M} + V(x) \tag{12}$$

the measured mass can be written as follows,

$$m^2 = M^2 + \frac{2M}{c^2} V \tag{13}$$

which coincides with $M$ in the free case $(V = 0)$ or in the nonrelativistic limit $(c \to \infty)$.

## 1.2. Covariant relativistic quantum mechanics

The procedure of quantization of the classical relativistic dynamics is a generalization of the quantization of Newtonian mechanics. The time $t$ is, in this case, a dynamical variable[2] just as the position $\vec{x}$, therefore, it is represented by a self-adjoint operator in the $L^2(R^4)$ Hilbert space, and enters in the wave function in the same way, which we denote as the covariant four vector $x^\mu$ (or, simply, $x$). The evolution of a quantum state, according to the Schrödinger-Stueckelberg equation, represented by the wave function, $\psi_\tau(x)$, is parameterized by $\tau$, as it is parameterized by $t$ in NRQM

$$i\hbar \frac{\partial \psi_\tau(x)}{\partial \tau} = K \psi_\tau(x) \tag{14}$$

The covariant commutation relations

$$[x^\mu, p^\nu] = i\hbar g^{\mu\nu}, \tag{15}$$

are assumed so that $p^0 = E/c$ (with spectrum $(-\infty, \infty)$) is the generator of translations in $t$, just as $\vec{p}$ is the generator of translations in $\vec{x}$.

The wave function $\psi_\tau(x)$ has the probabilistic interpretation that $|\psi_\tau(x)|^2$ is the probability density to find an event at the spacetime point $x^\mu$, and is normalized so that $\int d^4x |\psi_\tau(x)|^2 = 1$

---

[2] This property is responsible for the remarkable results of the Lindner experiment [13]

. In a more physical manner, $|\psi_\tau(x)|^2 d^4x$ is the probability of getting a signal (finding the particle) in the space region $d^3x$ around $\vec{x}$, and in the time interval $dt$ near $t$, that was generated by the particle in the historical time $\tau$.

## 1.3. Relative motion potential problem

The generalization of the isotropic interaction at a distance at equal time in NRQM for a two body system was studied by Horwitz and Piron [4]; it has the form $V(\rho)$, where

$$\rho^2 = (x_1 - x_2)^\mu (x_1 - x_2)_\mu \equiv (x_1 - x_2)^2 \tag{16}$$

represents the interaction between two space time events $x_1^\mu, x_2^\mu$ at equal $\tau$. In their paper [4] Horwitz and Arshansky showed that for invariant interval dependent potentials, the ground state wave function with support in a restricted $O(2,1)$ invariant subregion of the full spacelike region has a lower mass eigenvalue than for the full spacelike region (for which the Coulomb-like problem was solved by Cook [8] but gave a wrong spectrum), and yields the spectrum of the NR problem up to relativistic corrections. The evolution equation of two interacting events is of the form

$$i\frac{\partial}{\partial \tau}\psi_\tau(x_1, x_2) = K\psi_\tau(x_1, x_2) \tag{17}$$

where

$$K = \frac{p_1^2}{2M_1} + \frac{p_2^2}{2M_2} + V(\rho) \tag{18}$$

Separating the center of mass and the relative motion by defining

$$P^\mu = p_1^\mu + p_2^\mu \qquad X^\mu = \frac{M_1 x_1^\mu + M_2 x_2^\mu}{M_1 + M_2}$$

$$p^\mu = \frac{M_2 p_1^\mu - M_1 p_2^\mu}{M_1 + M_2} \qquad x^\mu = x_1^\mu - x_2^\mu \tag{19}$$

$$m = \frac{M_1 M_2}{M_1 + M_2} \qquad M = M_1 + M_2$$

Using these relations, the $K$ operator can then be written

$$K = \frac{P^2}{2M} + \frac{p^2}{2m} + V(\rho) = \frac{P^2}{2M} + K_{\text{Relative}} \qquad (20)$$

and the Stueckelberg-Schrödinger equation becomes

$$i\frac{\partial}{\partial \tau}\psi_\tau(X,x) = \left(\frac{P^2}{2M} + K_{\text{Relative}}\right)\psi_\tau(X,x) \qquad (21)$$

The center of mass part of $K$, has a continuous spectrum, but, for each $P$, $K_{\text{rel}}$ may have discrete spectrum, $K_a$. According to the Stueckelberg-Schrödinger equation, the wave function can then be written

$$\psi_\tau(P,x) = \exp\left(-i\frac{P^2}{2M}\tau\right)\exp(-iK_a\tau)\psi^{(a)}(P,x) \qquad (22)$$

For the relative motion we have an eigenvalue problem:

$$K_a \psi^{(a)}(P,x) = \left(-\frac{\partial^\mu \partial_\mu}{2m} + V(\rho)\right)\psi^{(a)}(P,x) \qquad (23)$$

The two body invariant mass squared (center of mass energy squared) is given by

$$s_a \equiv -P_a^2 = 2M(K_a - K) \qquad (24)$$

Choosing the subregion of Minkowski space (shown by Zmuidinas [6] to support complete sets of functions) that consists of the space time points external, in spacelike directions, to two hyperplanes tangent to the light cone that are oriented along the z axis (invariant under O(2,1)), Arshansky and Horwitz [4] showed that the solutions of this equation may have discrete spectrum (bound states) or continuous spectrum (scattering states).

We shall have special use for four-dimensional space time harmonic oscillator, of the form [9], [10]

$$V(\rho) = \frac{1}{2}m\omega^2 \rho^2 \qquad (25)$$

The $K_{\text{relative}}$ eigenvalue spectrum is given by [4]

$$K_{rl} = \hbar\omega\left(l + 2r + \frac{3}{2}\right) \qquad (26)$$

From (24), the CM energy spectrum can be written as (up to the second order)

$$E = \sqrt{-2Mc^2(K - K_{rl})} \simeq$$
$$\simeq \sqrt{-2Mc^2 K} + \sqrt{\frac{Mc^2}{2|K|}} \hbar\omega \left(l + 2r + \frac{3}{2}\right) - \sqrt{\frac{Mc^2}{8|K|^3}} \frac{\hbar^2\omega^2}{2} \left(l + 2r + \frac{3}{2}\right)^2 + ... \quad (27)$$

for small excitations relative to the total $K$. Assuming $K$ to be evaluated at the ionization point [4] for the free on-shell limit, $K = -Mc^2/2$, one obtains

$$E \simeq Mc^2 + \hbar\omega\left(l + 2r + \frac{3}{2}\right) - \frac{1}{2Mc^2}\hbar^2\omega^2\left(l + 2r + \frac{3}{2}\right)^2, \quad (28)$$

in agreement with the NR spectrum.

## 2. Motion of closed chain in Minkowski space

In this paper we study the system of $N$ identical particles in a closed chain interacting via an interval-dependent spring-like potential only with nearest neighbors.

First, we will develop the steady solution, in order to obtain the stable (equilibrium) shape of the chain. Second, a perturbation will be added which describes the oscillations around the equilibrium. We treat first the discrete chain and then the continuous limit.

The equation of motion for each of the particles of the chain is of the form

$$m \frac{d^2 q_{(n)}^\mu}{d\tau^2} = k\left(q_{(n+1)}^\mu - q_{(n)}^\mu\right) + k\left(q_{(n-1)}^\mu - q_{(n)}^\mu\right) \quad (29)$$

where we take all $m$ values equal, $n = 1, 2, ..., N$.

The covariant Hamiltonian then can be written as follows

$$K = \sum \frac{1}{2} m \dot{q}_{(n)\mu} \dot{q}_{(n)}^\mu + \sum \frac{1}{2} k \left(q_{(n+1)} - q_{(n)}\right)^\mu \left(q_{(n+1)} - q_{(n)}\right)_\mu \quad (30)$$

The Minkowski generalized "center of mass" (CM) is,

$$q_{CM}^\mu = \frac{\sum m_n q_{(n)}^\mu}{\sum m_n} = \frac{1}{N}\sum q_{(n)}^\mu \quad (31)$$

## 2.1. Steady solution

It is useful to define coordinates relative to the CM, $z_{(n)}^{\mu} = q_{(n)}^{\mu} - q_{CM}^{\mu}$, which satisfies, $\sum z_{(n)}^{\mu} = 0$. For an isolated system, $\ddot{q}_{CM}^{\mu} = 0$, the covariant Hamiltonian takes the form,

$$K = \frac{1}{2} M \dot{q}_{CM\mu} \dot{q}_{CM}^{\mu} + \sum \frac{1}{2} m \dot{z}_{(n)\mu} \dot{z}_{(n)}^{\mu} + \sum \frac{1}{2} k \left( z_{(n+1)} - z_{(n)} \right)_{\mu} \left( z_{(n+1)} - z_{(n)} \right)^{\mu} \quad (32)$$

We restrict all intervals to be spacelike; this restriction may be implemented by the parametrization (where we let $x, y, z, t$ be any of the relevant intervals)

$$\begin{aligned} x^0 &= \rho \sin\theta \sinh\beta & x^1 &= \rho \sin\theta \cos\phi \cosh\beta \\ x^2 &= \rho \sin\theta \sin\phi \cosh\beta & x^3 &= \rho \cos\theta \end{aligned} \quad (33)$$

The parameters $\theta, \phi, \beta, \rho$ completely span covariantly a 2+1 submanifold of the 3+1 Minkowski space, called the reduced Minkowski space (RMS). This submanifold was used for the support domain of the eigenfunctions of the two body bound state problem by Arshansky and Horwitz [4] which resulted in the correct Balmer series formulae in the nonrelativistic limit. The bound state wave functions vanish on the boundaries of this region, so there is no tunneling to the timelike regions.

A minimization of the Hamiltonian under the constraint of a constant chain length, $L$, determines that all the relative intervals between nearest neighbors have to be equal and constant: $\Delta z_{(j)} = L/N$. We seek a steady solution following simplifying assumptions:

First, assuming a constant interval between each particle and the CM, $\frac{d}{d\tau} z_{(n)\mu} z_{(n)}^{\mu} = 0$, the four-velocity may be written as $\frac{dz_{(n)}^{\mu}}{d\tau} = \Omega_{(n)\nu}^{\mu} z_{(n)}^{\nu}$.

Second, assuming a constant interval between each pair of nearest neighbor particles gives $\Omega_{(i)}^{\mu\nu} = \Omega_{(j)}^{\mu\nu} \equiv \Omega^{\mu\nu} \; \forall i, j$, which permits us to write the EOM as follows,

$$m \Omega^{\mu}_{\;\nu} \Omega^{\nu}_{\;\rho} z_{(i)}^{\rho} = k \left( z_{(i+1)}^{\mu} - z_{(i)}^{\mu} \right) + k \left( z_{(i-1)}^{\mu} - z_{(i)}^{\mu} \right) \quad (34)$$

Third, assuming only spacelike rotations i.e., $\left( \Omega^0_{\;i} = 0 \right)$, one can show that all the masses are in the same Minkowski hyperplane (the 3D space), forming a circular chain.

## 2.2. Effect of a geometric perturbation on the symmetric solution

A general perturbation, $z^{\mu}_{(n)} \to z^{\mu}_{(n)} + x^{\mu}_{(n)}$, to a steady solution results in the EOM

$$m \frac{d^2 x^{\mu}_{(n)}}{d\tau^2} = k \left[ \left( x^{\mu}_{(n+1)} - x^{\mu}_{(n)} \right) + \left( x^{\mu}_{(n-1)} - x^{\mu}_{(n)} \right) \right]$$

with Bloch waves solution,

$$x^{\mu}_{(n)}(\tau) = \sum_{l} x^{\mu}_{(n),l}(\tau) = \sqrt{\frac{2}{L}} \sum_{l}^{\infty} A^{\mu}_{l} \sin(W_l \tau) \sin(Q_l n \Delta L) \qquad (35)$$

where, $Q_l = \frac{2\pi}{L} l$, and, $W_l = \sqrt{\frac{2k}{m} \left( 1 - \cos\left( \frac{2\pi}{N} l \right) \right)}$

Inserting the solutions into the Hamiltonian, we get (it can be shown that the cross terms vanish)

$$K = \frac{M \dot{q}_{CM\mu} \dot{q}^{\mu}_{CM}}{2} + \frac{1}{2} \sum \left( m \dot{z}_{(n)\mu} \dot{z}^{\mu}_{(n)} + k \Delta z_{(n)\mu} \Delta z^{\mu}_{(n)} + m \dot{x}_{(n)\mu} \dot{x}^{\mu}_{(n)} + k \Delta x_{(n)\mu} \Delta x^{\mu}_{(n)} \right) \qquad (36)$$

The first term is a CM kinetic part, the second, is the steady solution kinetic term

$$K_{\text{Rotation}} = \sum \frac{1}{2} m \dot{z}_{(n)\mu} \dot{z}^{\mu}_{(n)} = \frac{1}{2} \Omega_{\mu}{}^{\alpha} \Theta_{\alpha}{}^{\beta} \Omega^{\mu}{}_{\beta}$$

where, $\Theta_{\alpha}{}^{\beta} \equiv \sum_{n=1}^{N} m z_{(n)\alpha} z^{\beta}_{(n)}$, is a generalization of "tensor of inertia" for the set of events in spacetime. The third part, $K_{\text{Tension}} = \sum \frac{1}{2} k \Delta z_{(n)\mu} \Delta z^{\mu}_{(n)} = \frac{1}{2} TL$, is the tension. The remaining part of the Hamiltonian describes the relative, perturbed, oscillations,

$$K_{\text{Oscillations}} = \sum \frac{1}{2} m \dot{x}_{(n)\mu} \dot{x}^{\mu}_{(n)} + \sum \frac{1}{2} k \Delta x_{(n)\mu} \Delta x^{\mu}_{(n)} \qquad (37)$$

Inserting these results into (36), the covariant Hamiltonian takes the form,

$$K = K_{CM} + K_{\text{Relative}} = \frac{M \dot{q}_{CM\mu} \dot{q}^{\mu}_{CM}}{2} + \frac{\Omega_{\mu}{}^{\alpha} \Theta_{\alpha}{}^{\beta} \Omega^{\mu}{}_{\beta}}{2} + \frac{TL}{2} + \frac{1}{2} \sum \left( m \dot{x}_{(n)\mu} \dot{x}^{\mu}_{(n)} + k \Delta x_{(n)\mu} \Delta x^{\mu}_{(n)} \right) \qquad (38)$$

Where,

$$K_{\text{Relative}} = K_{\text{Rotation}} + K_{\text{Tension}} + K_{\text{Oscillations}} \tag{39}$$

## 2.3. Continuous string

In the case of a continuous string, the four-tensor of inertia takes the form

$$\Theta_\alpha{}^\beta = \rho \int d\sigma z_\alpha(\sigma) z^\beta(\sigma) \tag{40}$$

where $\sigma$ is the parameter on the string, and assume, as for the discrete case, that all intervals, in the limit, are spacelike.

If we assume the whole string to be in $z^1 z^2$ plane, and to have the shape of a circle of perimeter $L$, rotating around the $z^3$ axis, the coordinates of the string would be, $z^0(\sigma) = z^3(\sigma) = 0$ and the rotation is described by $\Omega^{12} = \Omega_{12} = \Omega_z$. In that case, we get the familiar results for the tensor of inertia

$$\Theta_1^1 = \Theta_2^2 = \frac{MR^2}{2} = \frac{1}{2} \frac{ML^2}{(2\pi)^2} \qquad \Theta_1^2 = \Theta_2^1 = 0 \tag{41}$$

Inserting these into the rotation part of the Hamiltonian gives the known result,

$$K_{\text{Rotation}} = \frac{1}{2} \Omega_z^2 \frac{ML^2}{(2\pi)^2} \tag{42}$$

which, as can be shown, equals the tension part $K_{\text{Rotation}} = K_{\text{Tension}} = \frac{1}{2} TL$

## 3. The relative motion part of the covariant Hamiltonian

Now we turn to the motion around the equilibrium, $z_{(i)}^\mu$, represented by $x_{(i)}^\mu$. We start with the classical description and the representation in the mode space. Then, the system is quantized using the canonical commutation relations.

### 3.1. Oscillations of a string: classical treatment

In the continuous limit, where $\rho d\sigma = m$ and $\kappa d\sigma = \eta$, the Hamiltonian (37) takes the form,

$$K_{\text{Osc}} = \frac{1}{2} \int_0^L d\sigma \left( \rho \dot{x}^\mu \dot{x}_\mu + \eta \frac{dx^\mu}{d\sigma} \frac{dx_\mu}{d\sigma} \right) \tag{43}$$

where $x^\mu = x^\mu(\sigma,\tau) \in \mathbb{R}$, gives the EOM

$$\rho \frac{\partial^2 x^\mu}{\partial \tau^2} = \eta \frac{\partial^2 x^\mu}{\partial \sigma^2} \qquad (44)$$

We represent the solutions in terms of the Fourier expansion,

$$x^\mu(\sigma,\tau) = \sqrt{\frac{2}{L}} \sum_{n=1}^{\infty} \left[ A_n^\mu(\tau) \cos(Q_n \sigma) + B_n^\mu(\tau) \sin(Q_n \sigma) \right]$$

where $Q_n = n\frac{2\pi}{L}$, and $A^\mu, B^\mu$ are assumed to be spacelike as well, by our previous argument.

The dependence of the Fourier coefficients on $\tau$ is assumed to be of the form

$$A_n^\mu(\tau), B_n^\mu(\tau) \propto \cos(W_n \tau + \phi_n) \qquad (45)$$

The dispersion relation from the EOM is, $W_n = Q_n \sqrt{\frac{\eta}{\rho}} = n\frac{2\pi}{L}\sqrt{\frac{\eta}{\rho}} \equiv nW_0$. Using the inverse Fourier transform the coefficients may then be written as follows,

$$A_n^\mu(\tau) = \sqrt{\frac{2}{L}} \int_0^L d\sigma \cos(Q_n \sigma) x^\mu(\sigma,\tau), \quad B_n^\mu(\tau) = \sqrt{\frac{2}{L}} \int_0^L d\sigma \sin(Q_n \sigma) x^\mu(\sigma,\tau) \quad (46)$$

The canonical momentum density is of the form

$$p^\mu(\sigma,\tau) = \rho \dot{x}^\mu(\sigma,\tau) = \rho \sqrt{\frac{2}{L}} \sum_{n=0}^{\infty} \left[ \dot{A}_n^\mu(\tau) \cos(Q_n \sigma) + \dot{B}_n^\mu(\tau) \sin(Q_n \sigma) \right] \qquad (47)$$

We then have

$$\frac{\partial x^\mu(\sigma,\tau)}{\partial \tau} = \sqrt{\frac{2}{L}} \sum_{n=0}^{\infty} \left[ \dot{A}_n^\mu(\tau) \cos(Q_n \sigma) + \dot{B}_n^\mu(\tau) \sin(Q_n \sigma) \right]$$

$$\frac{\partial x^\mu(\sigma,\tau)}{\partial \sigma} = \sqrt{\frac{2}{L}} \sum_{n=0}^{\infty} Q_n \left[ -A_n^\mu(\tau) \sin(Q_n \sigma) + B_n^\mu(\tau) \cos(Q_n \sigma) \right]$$

The first term in the Hamiltonian takes the form

$$\int_0^L d\sigma \dot{x}^\mu \dot{x}_\mu = \frac{2}{L} \sum_{n,m=0}^\infty \int_0^L d\sigma \left[ \begin{array}{l} \dot{A}_n^\mu \dot{A}_{m\mu} \cos(k_n\sigma)\cos(k_m\sigma) + \dot{A}_n^\mu \dot{B}_{m\mu} \cos(k_n\sigma)\sin(k_m\sigma) + \\ \dot{B}_n^\mu \dot{A}_{m\mu} \sin(k_n\sigma)\cos(k_m\sigma) + \dot{B}_n^\mu \dot{B}_{m\mu} \sin(k_n\sigma)\sin(k_m\sigma) \end{array} \right] =$$

$$= \sum_{n=0}^\infty \left( \dot{A}_n^\mu \dot{A}_{n\mu} + \dot{B}_n^\mu \dot{B}_{n\mu} \right)$$

$$\int_0^L d\sigma \dot{x}^\mu \dot{x}_\mu = \sum_{n=0}^\infty \left( \dot{A}_n^\mu \dot{A}_{n\mu} + \dot{B}_n^\mu \dot{B}_{n\mu} \right) \tag{48}$$

The second term can be written as follows

$$\int_0^L d\sigma \frac{\partial x^\mu}{\partial \sigma} \frac{\partial x_\mu}{\partial \sigma} = \sum_{n=0}^\infty Q_n^2 \left( A_n^\mu A_{n\mu} + B_n^\mu B_{n\mu} \right) \tag{49}$$

The relative motion Hamiltonian then takes the form

$$K_{\text{Osc}} = \frac{1}{2} \sum_{n=0}^\infty \left( \rho \dot{A}_n^\mu \dot{A}_{n\mu} + \dot{B}_n^\mu \dot{B}_{n\mu} + \eta q_n^2 \left( A_n^\mu A_{n\mu} + B_n^\mu B_{n\mu} \right) \right) =$$

$$= \frac{1}{2} \rho \sum_{n=0}^\infty \left( \dot{A}_n^\mu \dot{A}_{n\mu} + \dot{B}_n^\mu \dot{B}_{n\mu} + W_n^2 \left( A_n^\mu A_{n\mu} + B_n^\mu B_{n\mu} \right) \right) \tag{50}$$

Since it is diagonalized in the mode space, it can be written as a sum, $K_{\text{Osc}} = \sum_{n=0}^\infty K_n^{\text{Osc}}$, where

$$K_n^{\text{Osc}} = \frac{1}{2} \rho \left[ \left( \dot{A}_n^\mu \dot{A}_{n\mu} + \dot{B}_n^\mu \dot{B}_{n\mu} \right) + \omega_n^2 \left( A_n^\mu A_{n\mu} + B_n^\mu B_{n\mu} \right) \right] \tag{51}$$

### 3.2. Oscillations of a string: quantum treatment

The variables $x^\mu, p^\mu$ (and therefore $A^\mu, B^\mu$) are now operators. We will look for eigenfunctions:

$$K_{\text{Osc}} \varphi^\lambda = k^\lambda \varphi^\lambda \tag{52}$$

We assume commutation relations

$$\left[ x^\mu(\sigma,\tau), p^\nu(\sigma',\tau) \right] = i\delta(\sigma-\sigma') g^{\mu\nu} \tag{53}$$

Using the definition of the canonical momentum density, this can be written as

$$\left[x^{\mu}(\sigma,\tau),\dot{x}^{\nu}(\sigma',\tau)\right]=\frac{1}{\rho}i\delta(\sigma-\sigma')g^{\mu\nu} \tag{54}$$

So that

$$\left[A_n^{\mu}(\tau),\dot{A}_m^{\nu}(\tau)\right]=\frac{2}{L}\int_0^L d\sigma\cos(Q_n\sigma)\cos(Q_m\sigma')\left[x^{\mu},\dot{x}^{\mu}\right]=\frac{1}{\rho}i\delta_{nm}g^{\mu\nu} \tag{55}$$

We now define a momentum operator conjugate to $A_n^{\mu}$ by

$$\pi_{n,A}^{\mu}=\rho\dot{A}_n^{\mu}=-i\frac{\partial}{\partial A_{n\mu}} \tag{56}$$

so that

$$\left[A_n^{\mu}(\tau),\pi_{mA}^{\nu}(\tau)\right]=i\delta_{nm}g^{\mu\nu} \qquad \left[B_n^{\mu}(\tau),\pi_{mB}^{\nu}(\tau)\right]=i\delta_{nm}g^{\mu\nu}$$

Then $K_n$ takes the form

$$K_n^{\text{Osc}}=\left[\frac{1}{2\rho}\left(\pi_{nA}^{\mu}\pi_{nA\mu}+\pi_{nB}^{\mu}\pi_{nB\mu}\right)+\frac{1}{2}\rho W_n^2\left(A_n^{\mu}A_{n\mu}+B_n^{\mu}B_{n\mu}\right)\right] \tag{57}$$

This appears as a set of two body 4D oscillators which may be solved by the method of Arshansky and Horwitz in 1989 [4].

The eigenfunctions satisfy

$$\left[\frac{1}{2\rho}\left(\pi_{nA}^{\mu}\pi_{nA\mu}+\pi_{nB}^{\mu}\pi_{nB\mu}\right)+\frac{1}{2}\rho\omega_n^2\left(A_n^{\mu}A_{n\mu}+B_n^{\mu}B_{n\mu}\right)\right]\varphi_n^{\lambda(n)}\left(A_n^{\mu},B_n^{\mu}\right)=$$
$$=k_n^{\lambda(n)}\varphi_n^{\lambda(n)}\left(A_n^{\mu},B_n^{\mu}\right) \tag{58}$$

Separating variables, we get

$$\varphi_n^{\lambda(n)}\left(A_n^{\mu},B_n^{\mu}\right)=\varphi_n^{\lambda_A(n)}\left(A_n^{\mu}\right)\varphi_n^{\lambda_B(n)}\left(B_n^{\mu}\right) \tag{59}$$

where each function satisfies

$$K_{n(A)}^{\text{Osc}}\varphi_n^{\lambda_A(n)}\left(A_n^\mu\right) = \left[\frac{1}{2\rho}\pi_{nA}^\mu \pi_{nA\mu} + \frac{\rho}{2}W_n^2 A_n^\mu A_{n\mu}\right]\varphi_n^{\lambda_A(n)}\left(A_n^\mu\right) = k_n^{\lambda_A(n)}\varphi_n^{\lambda_A(n)}\left(A_n^\mu\right)$$

$$K_{n(B)}^{\text{Osc}}\varphi_n^{\lambda_B(n)}\left(B_n^\mu\right) = \left[\frac{1}{2\rho}\pi_{nB}^\mu \pi_{nB\mu} + \frac{\rho}{2}W_n^2 B_n^\mu B_{n\mu}\right]\varphi_n^{\lambda_B(n)}\left(B_n^\mu\right) = k_n^{\lambda_B(n)}\varphi_n^{\lambda_B(n)}\left(B_n^\mu\right)$$

The total eigenvalue is of the form

$$k_n^{\lambda(n)} = k_n^{\lambda(n,A)} + k_n^{\lambda(n,B)} \tag{60}$$

As said above in the introduction section, this problem has been solved [4] in the restricted Minkowski space (RMS) for the "coordinates" $A_\mu, B_\mu$. It has discrete spectrum, labelled by $\lambda_A(n), \lambda_B(n)$ sets of 4 quantum numbers for the solutions

$$\varphi^\lambda(A) = R_r^l(\rho)\Theta_l^p(\theta)B_{mp}(\beta)\Phi_m(\phi) \tag{61}$$

The eigenvalue set for a given wave function is given by

$$\lambda = (l, r, p, m) \tag{62}$$

The general wave function, including all the modes, is then given by

$$\varphi^{\{\lambda\}}(A_1, A_2, \ldots B_1, B_2, \ldots) = \prod_{n=1}^{\infty} \varphi_n^{\lambda_A(n)}(A_n)\varphi_n^{\lambda_B(n)}(B_n) \tag{63}$$

where we defined the eigenvalues set as

$$\{\lambda\} \equiv \lambda_A(1), \lambda_B(1), \lambda_A(2), \lambda_B(2), \ldots = \{\lambda_A, \lambda_B\} \tag{64}$$

## 4. Superposition

The general solution is given by the superposition

$$\psi(A_1 \ldots B_1 \ldots) = \sum_{\{\lambda\}} C(\{\lambda\})\varphi^{\{\lambda\}} = \sum_{\{\lambda\}} C(\{\lambda\})\prod_{n=1}^{\infty} \varphi_n^{\lambda_A(n)}(A_n)\varphi_n^{\lambda_B(n)}(B_n) \tag{65}$$

Just as $|\varphi(\vec{x})|^2 d^3x$ is a probability "to find" a particle in position $x$ in the volume $d^3x$ in the NR theory, so, $|\varphi_n^{\lambda_A}(A_n)|^2 d^4A_n$ is the probability "to find" the Fourier coefficient $A_n^\mu$ (with set of quantum numbers $\lambda_A$) in the region $d^4A$, with the amplitude playing the role of canonical coordinates of the oscillator, $\sum |C(\{\lambda\})|^2 = 1$. The Hilbert space is defined in terms of the normalization.

$$\left\langle \varphi_n^{\lambda_A(n)} \middle| \varphi_n^{\lambda_A(n)} \right\rangle = \int d^4 A_n \left( \varphi_n^{\lambda_A(n)}(A_n) \right)^* \varphi_n^{\lambda_A(n)}(A_n) = 1 \quad ^3 \tag{66}$$

i.e.,

$$\int d^4 A_n |A_n\rangle\langle A_n| = 1$$

Since each of the functions is normalized, so is a product of them

$$\int \prod_{i=1}^{\infty} d^4 A_i d^4 B_i \left[ \left( \varphi_i^{\lambda_A(i)}(A_i) \right)^* \varphi_i^{\lambda_A(i)}(A_i) \right] \left[ \left( \varphi_i^{\lambda_B(i)}(B_i) \right)^* \varphi_i^{\lambda_B(i)}(B_i) \right] = 1$$

$$\left\langle \varphi^{\{\lambda\}} \middle| \varphi^{\{\lambda\}} \right\rangle = \int \prod_{i=1}^{\infty} d^4 A_i d^4 B_i \left| \varphi_i^{\lambda_A(i)}(A_i) \right|^2 \left| \varphi_i^{\lambda_B(i)}(B_i) \right|^2 = 1$$

$$\sum_n \int d^4 A_n |A_n\rangle\langle A_n| = 1$$

The superposition state is also normalized

$$\langle \psi | \psi \rangle = \int \prod_i d^4 A_i d^4 B_i \left| \psi(\{A,B\}) \right|^2 =$$
$$= \int \prod_i d^4 A_i d^4 B_i \sum |C(\{\lambda\})|^2 \left| \varphi_i^{\lambda_A(i)}(A_i) \right|^2 \left| \varphi_i^{\lambda_B(i)}(B_i) \right|^2 = \sum |C(\{\lambda\})|^2 = 1 \tag{67}$$

We calculate $\left\langle A_m^\mu \right\rangle$ in the following.

For a single wavefunction $\varphi^\lambda$

$$\left\langle A_m^\mu \right\rangle = \left\langle \varphi^\lambda \middle| A_m^\mu \middle| \varphi^\lambda \right\rangle = \int A_m^\mu \prod_i d^4 A_i d^4 B_i \left| \varphi_i^{\lambda_A(i)}(A_i) \right|^2 \left| \varphi_i^{\lambda_B(i)}(B_i) \right|^2 =$$
$$= \int A_m^\mu d^4 A_m \left| \varphi_m^{\lambda_A(m)}(A_m) \right|^2 \tag{68}$$

---

[3] Note that we are dealing with a Hilbert bundle, as for the usual quantization of the electromagnetic field.

For superposition states, we have

$$\langle A_m^\mu \rangle = \langle \psi | A_m^\mu | \psi \rangle = \int A_m^\mu \prod_i d^4 A_i d^4 B_i \, |\psi(\{A,B\})|^2 =$$
$$= \int A_m^\mu \prod_i d^4 A_i d^4 B_i \sum |C(\{\lambda\})|^2 \left|\varphi_i^{\lambda_A(i)}(A_i)\right|^2 \left|\varphi_i^{\lambda_B(i)}(B_i)\right|^2 = \quad (69)$$
$$= \sum |C(\{\lambda\})|^2 \int d^4 A_m A_m^\mu \left|\varphi_m^{\lambda_A(m)}(A_m)\right|^2$$

Hence,

$$\langle x^\mu(\sigma) \rangle = \sqrt{\frac{2}{L}} \sum_{n=1}^\infty \left[ \langle A_n^\mu \rangle \cos(q_n \sigma) + \langle B_n^\mu \rangle \sin(q_n \sigma) \right] \quad (70)$$

The state where all the "oscillators" are in the lowest level of spectrum we call the vacuum of the excitations or the ground state

$$|0\rangle \equiv \prod_{i=1}^\infty \varphi_i^{\lambda_A(i)=GS}(A_i) \varphi_i^{\lambda_B(i)=GS}(B_i) \quad (71)$$

So, in writing the string state, we may consider only the excited oscillators:

$$|\lambda_A(n_1),...,\lambda_A(n_N),\lambda_B(m_1),...,\lambda_B(m_M)\rangle \equiv$$
$$\equiv \varphi_{n_1}^{\lambda_A(n_1)}(A_{n_1})...\varphi_{n_N}^{\lambda_A(n_N)}(A_{n_N}) \varphi_{m_1}^{\lambda_B(m_1)}(B_{m_1})...\varphi_{m_M}^{\lambda_B(m_M)}(B_{m_M}) \prod_{\substack{i \neq n_1,...,n_N, \\ m_1,...,m_M}}^{\infty} \varphi_i^{\lambda_A(i)=GS}(A_i) \varphi_i^{\lambda_B(i)=GS}(B_i)$$

We remark that this procedure achieves "second quantization" without putting the Fourier coefficients into correspondence with annihilation-creation operators, but uses the wave functions which are solutions of the differential equations with the coefficients as variables to construct the many mode functions (through the direct product). This avoids the problem of conflict between covariance and positive norm.

## 5. The spectrum

The Hamiltonian operator acting on a general wave function gives

$$K_n^{Osc} \varphi^{\{\lambda\}} = \left( K_{n(A)}^{Osc} + K_{n(B)}^{Osc} \right) \prod_{n=1}^\infty \varphi_n^{\lambda_A(n)}(A_n) \varphi_n^{\lambda_B(n)}(B_n) =$$
$$= \sum_{n=1}^\infty \left( k^{\lambda_A(n)} + k^{\lambda_B(n)} \right) \prod_{n=1}^\infty \varphi_n^{\lambda_A(n)}(A_n) \varphi_n^{\lambda_B(n)}(B_n) = k^{\{\lambda\}} \varphi^{\{\lambda\}} \quad (72)$$

For each mode $n$ $\left(\varphi_n^{\lambda_A(n)}\left(A_n^\mu\right), \varphi_n^{\lambda_B(n)}\left(B_n^\mu\right)\right)$ the eigenvalues are known [4]

$$k_n^\lambda = k_n^{lr} = \hbar W_n \left(l_n + 2r_n + \frac{3}{2}\right) \tag{73}$$

In the formulation we have given in the RMS, the $l_n$ are the physical angular momenta (see (24) and (26) and page 86 of ref [3]). This result is consistent with the observed Regge relation between mass squared and angular momentum for each value of the principal quantum numbers $r_n$ labelling families of trajectories. We remark that the eigenvalues in (73) are the result of properties of the special functions (Laguerre functions, [3]) that are solutions of the eigenvalue equations.

The general wave function eigenvalue hence takes the form

$$k^{\{\lambda\}} = \sum_{n=1}^\infty \left(k^{\lambda_A(n)} + k^{\lambda_B(n)}\right) \tag{74}$$

In terms of wave function indices, it can be written as follows

$$\begin{aligned} K_{\text{Osc}}\left(\{l_n^A, r_n^A, l_n^B, r_n^B\}\right) &= k\left(\{l_n^A, r_n^A, l_n^B, r_n^B\}\right) = \\ &= \sum_n \hbar W_n \left(l_n^A + 2r_n^A + l_n^B + 2r_n^B + 3\right) \end{aligned} \tag{75}$$

The ground state eigenvalue would be

$$K_{\text{Osc}}^{GS} = 3\sum_n \hbar W_n = 3\sum_n \hbar W_0 n = 3\hbar W_0 \sum_n n \tag{76}$$

The zero-point energy, as for the usual classical oscillator, diverges, and we may cancel this with a phase.

Using (20), the invariant mass squared of the system (center of mass energy squared) can be written as follows

$$s \equiv -P^\mu P_\mu = 2M \left(K_{\text{Relative}} - K\right) \tag{77}$$

where $M = \rho L$. Setting $K = -\dfrac{Mc^2}{2}$

$$s \equiv -P^\mu P_\mu = M^2 c^2 + 2M K_{\text{Relative}} \tag{78}$$

Hence, the mass squared takes the form

$$m^2 = M^2 + \frac{2M}{c^2} K_{Rel} = M^2 \left(1 + \frac{2}{Mc^2} K_{Relative}\right) \quad (79)$$

and, using (39), and (75) it can be written as (the zero point contribution in (80) and (82) may be deleted; we leave it here to display the effective dimensionality of the bound state problem)

$$m^2 = M^2 \left[1 + \frac{2}{Mc^2} TL + \frac{2}{Mc^2} \sum_n \hbar W_n \left(l_n^A + 2r_n^A + l_n^B + 2r_n^B + 3\right)\right] \quad (80)$$

For the CM frame, where, $E^2 = m^2 c^4$, the energy spectrum is

$$E = Mc^2 \sqrt{1 + \frac{2}{Mc^2} K_{Rel}} \quad (81)$$

In the nonrelativistic limit, it may be approximated, to the first order, by

$$E \simeq Mc^2 + K_{Rel} = Mc^2 + TL + \sum_n \hbar W_n \left(l_n^A + 2r_n^A + l_n^B + 2r_n^B + 3\right) \quad (82)$$

## 6. Summary and discussion

The covariant description of the relativistic closed string was carried out in the framework of Stueckelberg-Horwitz-Piron formalism. It was shown that the string motion can be separated into three independent parts: center of mass motion, rotation, and relative oscillations. The oscillations part is quantized, while, the other parts are treated classically. The rotational part of the Hamiltonian is obtained from a demand for a steady solution, which showed that the string's equilibrium configuration is found in the same hyperplane (a spacelike hyperplane for consistency with the nonrelativistic limit). Therefore, the system can be transformed to some Lorentz frame where the string is embedded in the usual **R**$^3$. Assuming equal spacing and constant $\rho$ to the center, the string's steady solution's shape turns out to be a rotating ring. The remaining part of the Hamiltonian describes the relative oscillations around the equilibrium (steady solution). The Fourier expansion of the oscillation Hamiltonian behaves like a set of independent relativistic harmonic oscillators representing excitations of the string's modes. By defining the mode variables as coordinates (second quantization) and applying the methods of Arshansky and Horwitz [4], we find the spectrum of the oscillatory motion of the string. Sections 3 and 4 constitute an essentially new treatment for the covariant quantization of an extended system.

We finally remark that Regge theory applied to high energy scattering implies that there are families of particles with mass-squared (bosons) increasing, to some approximation, (experimentally) linearly rising with their intrinsic angular momentum. This result is consistent with string theory (e.g. Polchinski [11]) based on an essentially geometric picture of the string. Our approach takes a more dynamical point of view, for which the string excitations are the result of local oscillator interactions.

Current string models use a reparameterization invariant Lagrangian, while the present model uses a Hamiltonian formalism[4], providing a wider basis for application to dynamical models in terms of both particles and quantum field theory. In this framework there is no need for additional dimensions. These topics will be the subject of future research.

---

[4]Frastai and Horwitz [16] have pointed out that, as seen from the path integral approach of Schwinger and Feynman, that the off-shell theory does not have a Hamiltonian representation if it is reparametrization invariant. As a simple example, consider a reparametrization invariant Lagrangian of the form $\int \sqrt{\dot{q}^\mu \dot{q}_\mu} ds$ (the dots are differentiation by s). Then $p^\mu$ is proportional to $\dfrac{\dot{q}^\mu}{\sqrt{\dot{q}^\nu \dot{q}_\nu}}$, and $p^\mu$ is then constant, a necessarily on shell picture.